\documentclass[twocolumn]{aastex62}

\usepackage{amsmath}
\usepackage{booktabs}
\usepackage{dcolumn}
\usepackage{gensymb}
\usepackage{hyperref}
\usepackage{color}

\shorttitle{TRGB Distance to the Fireworks Galaxy}
\shortauthors{Anand, Rizzi, $\&$ Tully}

\begin{document}
\title{A Robust Tip of the Red Giant Branch Distance to the Fireworks Galaxy (NGC 6946)}
\author{Gagandeep S. Anand}
\affiliation{Institute for Astronomy, University of Hawaii, 2680 Woodlawn Drive, Honolulu, HI 96822, USA}
\author{Luca Rizzi}
\affiliation{W. M. Keck Observatory, 65-1120 Mamalahoa Hwy, Kamuela, HI 96743, USA}
\author{R. Brent Tully}
\affiliation{Institute for Astronomy, University of Hawaii, 2680 Woodlawn Drive, Honolulu, HI 96822, USA}

\begin{abstract}
Archival \textit{HST} data taken in F606W+F814W of two different fields in the outer regions of NGC 6946 is used to measure a tip of the red giant branch (TRGB) distance to the galaxy. We employ a Bayesian maximum-likelihood modeling method that incorporates the completeness of the photometry, and allows us to model the luminosity function of the RGB population. Our two fields provide us with distances of 7.74 $\pm$ 0.42 Mpc and 7.69 $\pm$ 0.50 Mpc, respectively. Our final distance of 7.72 $\pm$ 0.32 Mpc is higher than most values published previously in the literature. This has important implications for supernova measurements, as NGC 6946 is host to the most observed supernovae (10) of any galaxy to date. We find that the supernovae in NGC 6946 are on average $\sim$ 2.3 times more luminous than previous estimates. Our distance gives NGC 6946 a peculiar velocity of $v_{pec}$ = $-229$ $\pm$ $29$ km/s in the Local Sheet frame. This velocity is the projected component of a substantial negative SGZ motion, indicating a coherent motion along with the other members of the M101 wall toward the plane of the Local Sheet. The M101 wall, a spur off the Local Sheet into the Local Void, is experiencing the expansion of the Local Void.
\end{abstract}


\section{Introduction} \label{intro}

On the largest scales, the Universe is homogeneous and isotropic. But on smaller scales, there is plenty of organized structure. This structure takes the shape of features such as cosmic voids, filaments, and superclusters. The precise nature of these structures depends greatly on the underlying cosmological models and parameters. Investigating this organized structure in detail provides us with important observational constraints, and may allow us to discern between separate models. 

In order to determine the precise nature of nearby structures, we need to obtain accurate distances to the galaxies that inhabit them. This exercise can be challenging. In addition to the velocity due to the Hubble flow ($H_{0}d$), a galaxy also has an intrinsic velocity caused by gravitational interactions with other galaxies. This peculiar velocity ($v_{pec}$) is an additional component of the observed velocity, such that

\begin{equation}
v_{obs} = v_{pec} + H_{0} d
\end{equation}

In the local Universe, the peculiar velocity is much more likely to be a significant component of the total observed velocity, where the velocity due to the Hubble flow is lower. This circumstance can be turned to advantage. Nearby, individual peculiar velocities can be determined with modest uncertainties through accurate measured distances.

\begin{figure*}
\centering
\figurenum{1}
\epsscale{1.15}
\plottwo{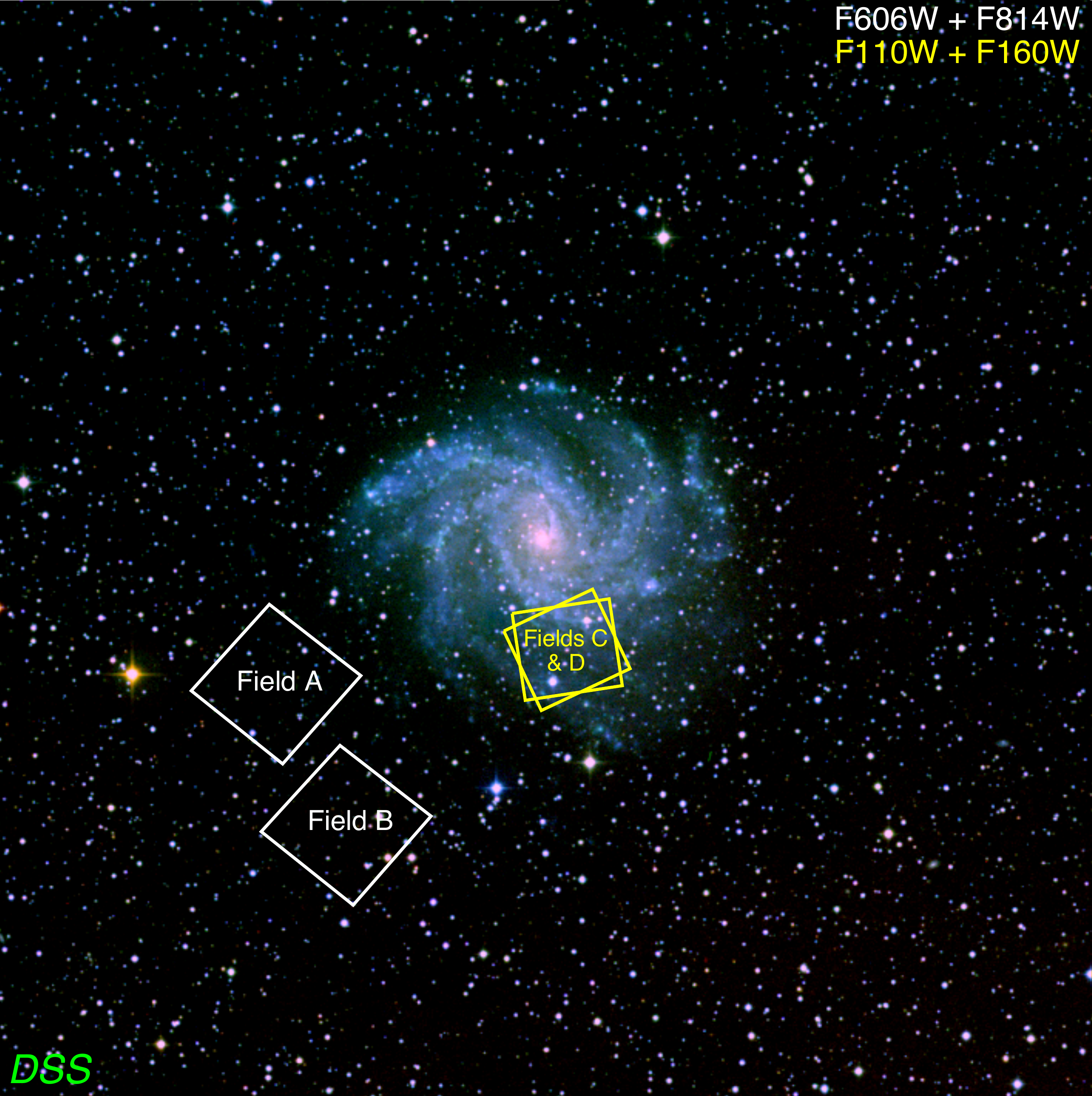}{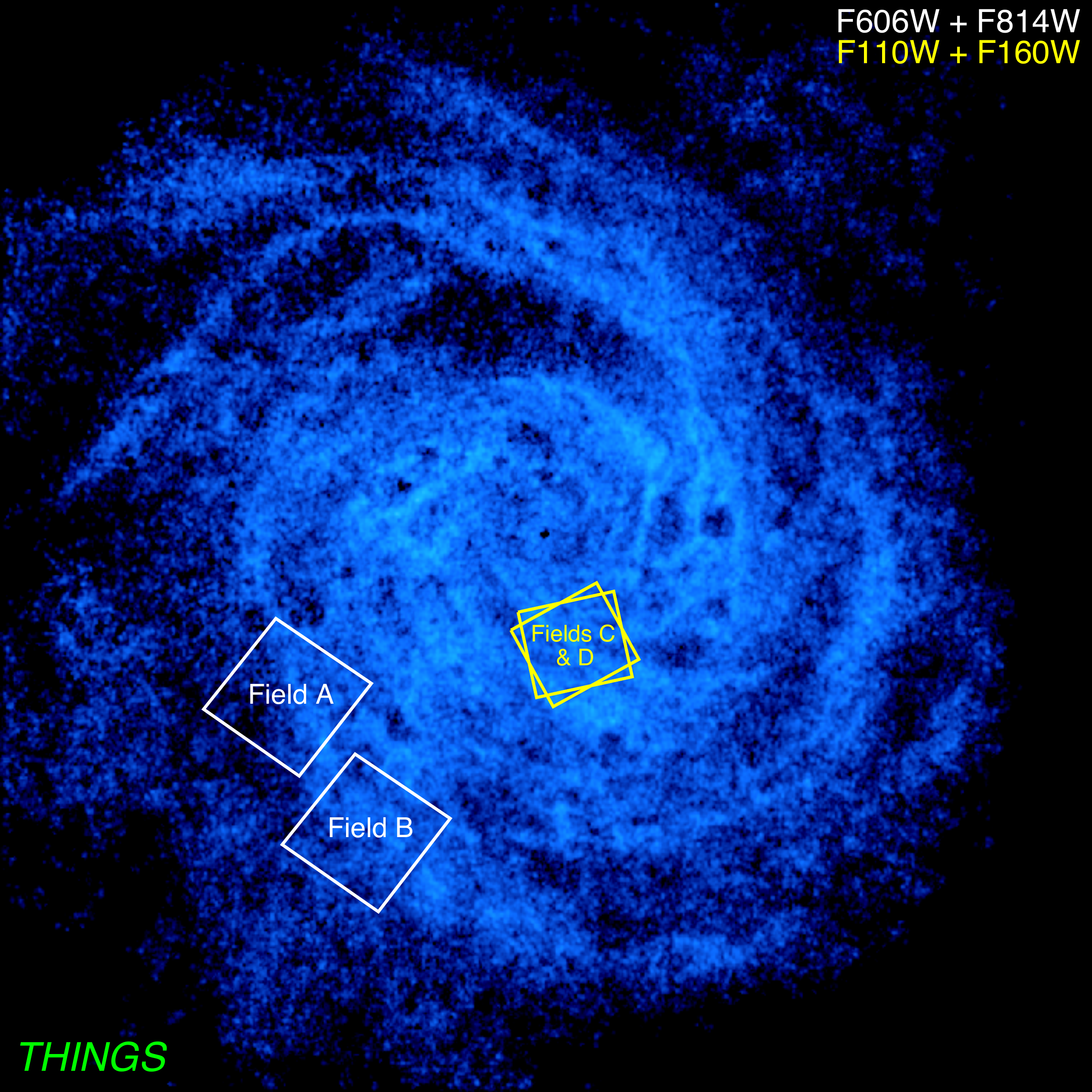}
\caption{The left panel of this figure shows our four HST fields overlaid onto a color image of NGC 6946 created from the Digitized Sky Survey (red from POSS2/UKSTU IR, green from POSS2/UKSTU Red, and blue from POSS2/UKSTU Blue). The spatial extent of the image is $\sim$26' x 26'. The right panel shows the fields overlaid on an HI map from THINGS. The key in the upper right hand corner show the HST filters of the observations. Note that the radial extent of the HI emission is greater than that of the optical from the DSS - although Fields A $\&$ B do not appear to correspond with spiral arms in the optical, the radio image shows that they do.}
\label{sidebyside}
\end{figure*}

\subsection{Tip of the Red Giant Branch }

One method that is becoming increasingly popular to determine distances to nearby galaxies uses the brightest red giant stars in a galaxy as a standard candle. This is known as the tip of the red giant branch (TRGB) method \citep{dacosta1990, ogTRGB}. Stars evolving off the main sequence and climbing the red giant branch will eventually reach a point where the temperature is hot enough that the degenerate helium core can fuse helium via the triple-$\alpha$ process. This runaway process (the helium flash) is followed by the transition to the horizontal branch. Since the maximum mass of the helium core attained on the RGB is a constant, this leads to the brightest RGB stars acting as a standard candle. The magnitude of this tip has been well-calibrated, is fairly insensitive to metallicity, (especially in the $I$-band, where $M_{I}$ $\sim$ $-$4.05), and can be determined with high precision ($\sim5\%$) when the photometry is of high spatial resolution, such as that from \textit{HST} \citep{rizzi2007}. The TRGB method also has several advantages when compared to distances obtained from Cepheid variables. There is no requirement of a temporal baseline, and there are fewer issues with extinction within the host galaxy (population II stars associated with the TRGB measurements lie mostly outside of the thin disk of spiral galaxies).

Given its standard candle nature, once we know the apparent magnitude of the TRGB, we can correct for extinction and minor metallicity effects, and obtain a distance to the galaxy. 

\subsection{NGC 6946}

NGC 6946 is a face-on starburst galaxy located 12$\degree$ off the plane of the Milky Way. There are currently ten recorded supernovae (no type Ia) that have occurred within the galaxy\footnote{\url{https://sne.space/}}, giving rise to the nickname ``The Fireworks Galaxy''. Current distance estimates range from 4.0 Mpc to 12.7 Mpc in NED\footnote{\url{https://ned.ipac.caltech.edu/}}. 

Unfortunately, the distances obtained from supernovae have a large spread, spanning the full range of distances noted above. A good distance to this galaxy will allow us to better understand the flows of galaxies in the local Universe, and the way structure forms. In this case, NGC 6946 is located along a long filament (the M101 wall) that protrudes off of the Local Sheet into the Local Void \citep{muller2017}. An accurate distance will allow us to pin down its relationship to the wall, as well as determine the velocity of recession from the Local Void. A good distance also improves constraints on supernova models by providing absolute luminosities for the ten supernovae that have occurred within this galaxy.

In this paper, we use the TRGB method on multiple fields of archival \textit{HST} imaging to find a single, robust distance to NGC 6946. We then discuss implications for large-scale structure, as well as supernova science. Errors presented represent 1$\sigma$ confidence intervals, unless otherwise specified. 

\begin{table*}[]
\centering
\begin{tabular}{|c|c|c|c|c|}
\hline
\textbf{Proposal ID} & \textbf{PI} & \textbf{Photometric System} & \textbf{Filters/Total Observation Times}     & \textbf{Field Name} \\ \hline
12331                & C. Kochanek & WFC3/IR                     & F110W (1598s), F160W (4196s)                 & Field C             \\ \hline
12450                & C. Kochanek & WFC3/IR                     & F110W (1498s),  F160W (1398s)                & Field D             \\ \hline
14786                & B. Williams & WFC3/UVIS                   & F606W (5470s + 5470s), F814W (5548s + 5538s) & Fields A + B        \\ \hline
\end{tabular}
\caption{This table identifies the archival observations used in our determining of a TRGB distance to NGC 6946, along with some key parameters. Note that we were only able to use Fields A $\&$ B for the final determination, as the NIR fields do not reach the appropriate depth.}
\label{obsTable}
\end{table*}

% NGC 6946 is one of few large galaxies within 10 Mpc that has lacked a robust published distance. 

\section{Observations} \label{obs}

We searched for archival \textit{HST} data of NGC 6946 on the Mikulski Archive for Space Telescopes (MAST)\footnote{\url{https://archive.stsci.edu/hst/}}. After combing through all the publicly available data, we found three proposals that we believed contained observations with the appropriate filters and depths required to achieve a TRGB measurement (though later we find that the NIR observations are too shallow). These are \textit{HST} proposals 12331 $\&$ 12450 \citep{12331,12450}, $\&$ 14786 \citep{14786}. The first two proposals each contain observations of a field (with large overlaps, but not identical) in the near-infrared (F110W and F160W). The latter proposal contains observations in the optical (F606W and F814W) in two separate, but nearly adjacent fields. In order to obtain a better sense of the observations, we have overlaid the observed fields onto optical and radio images (Figure \ref{sidebyside}) obtained from the Digitized Sky Survey (DSS) and The HI Nearby Galaxies Survey, THINGS \citep{THINGS}. We see that the HI emission from the galaxy is much further extended than the optical emission, resulting in a non-negligible quantity of blue stars in Fields A $\&$ B, despite their radial extent from the center of the galaxy. A further summary of the observations is provided in Table \ref{obsTable}.

\section{Data Reduction $\&$ TRGB Magnitudes} \label{dataTRGB}
Our overall reduction process is based on \cite{MLM}, which has been used as the basis for many recent TRGB papers from our wider collaboration \citep{K2015,rizzi2017,L2018,Anand2018}. In this section, we give a brief overview of our reduction procedure. We also highlight some of the challenges we faced in the process determining a TRGB magnitude for NGC 6946 (continued in \S 4).

\subsection{Photometry}
We use the latest version (v2.0, August 2017 release) of the PSF photometry software DOLPHOT \citep{dolphot,dolphot2} to perform photometry on our four fields. DOLPHOT is specifically built to handle \textit{HST} photometry, and includes PSFs of all ACS and WFC3 filters. In addition to the required stellar photometry, we use DOLPHOT to perform completeness simulations on our fields, which involves inserting and attempting to recover artificial stars. 

\subsection{Spatial Selection}
\begin{figure*}
\figurenum{2}
\epsscale{1.1}
\plottwo{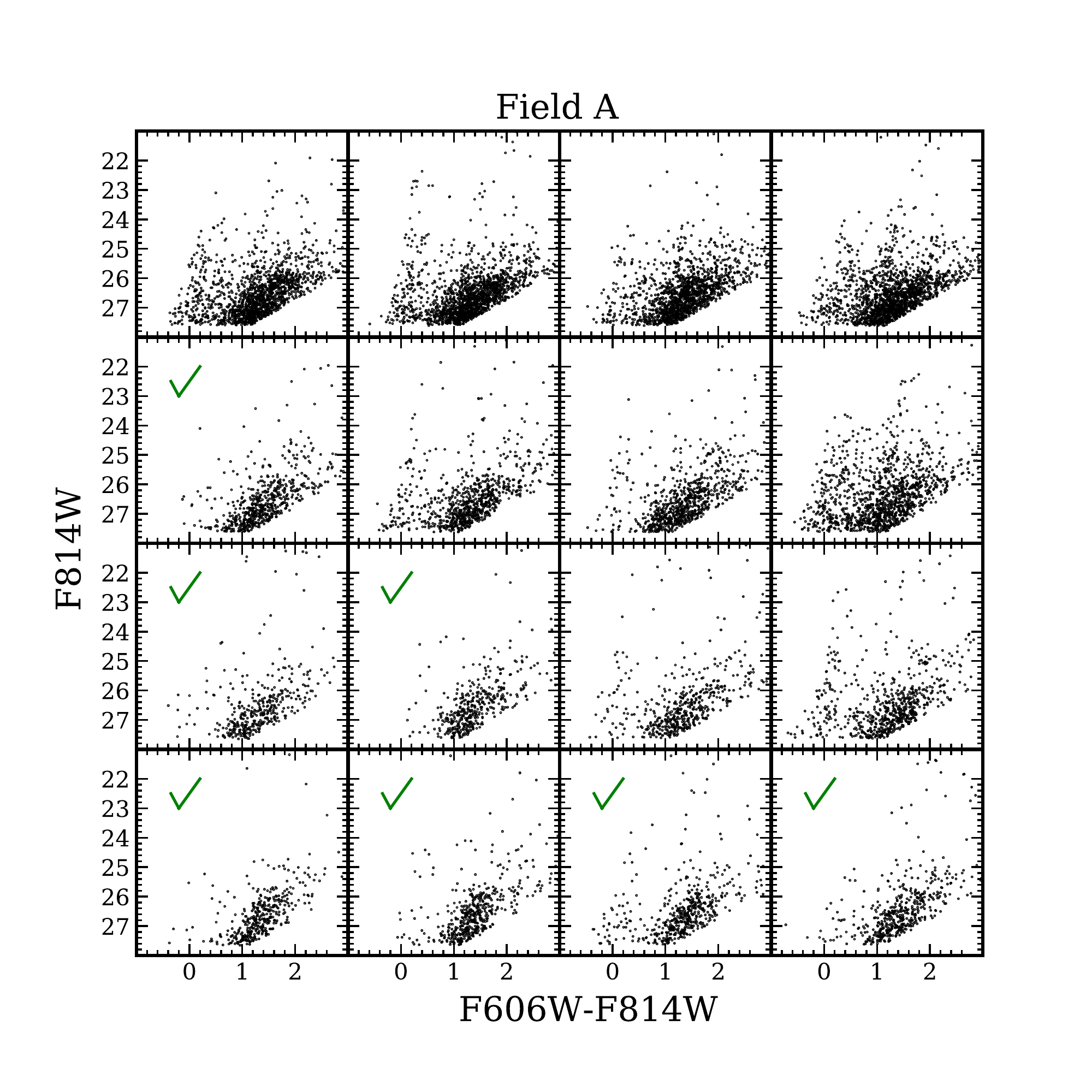}{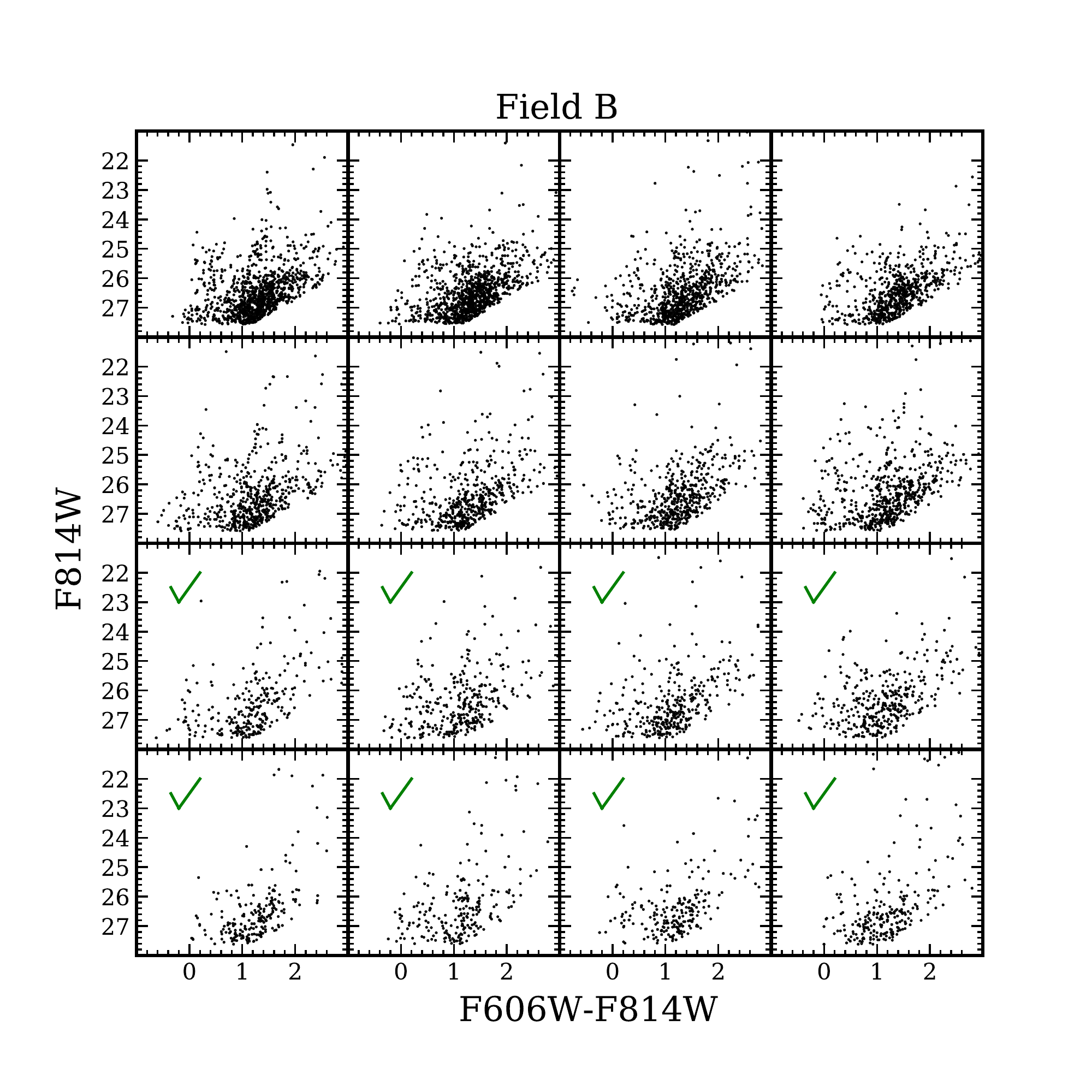}
\caption{Color magnitude diagrams for 16 subfields in each optical field. We mark the ones with lower overall levels of Population I stars with check marks. These are the ones that are used for the final determination of the magnitude of the TRGB. This procedure was more fruitful for Field A, as part of the field lies in a region with lower HI content (Figure \ref{sidebyside}).}
\label{xyplots}
\end{figure*}

To obtain the best possible estimate of the TRGB, we divided up the field in 16 boxes and plotted individual color-magnitude diagrams for each (Figure \ref{xyplots}). We looked for subfields with higher overall levels of Population I stars, and excluded them from our analysis to cut down on contamination from AGB stars that may be internally reddened to the extent that they appear as RGB stars. For both Fields A $\&$ B, we ended up using half of the available field. This procedure was more fruitful for Field A, as part of the field lies in a region with noticeably lower HI content (Figure \ref{sidebyside}). This situation is true for Field B as well, but to a lesser extent.

\subsection{Measuring the TRGB}
With the photometry and completeness simulations in hand, we proceed to measure the apparent magnitude of the TRGB. The method we use is based on a maximum likelihood algorithm implemented and tested heavily by \cite{MLM}. They introduce a theoretical stellar luminosity function of the form
\begin{equation}
\psi = \left\{\begin{matrix}10^{a(m-m_{TRGB})+b}, m-m_{TRGB} \geq 0 \\ 
\\
10^{c(m-m_{TRGB})}, \ \ \ m-m_{TRGB} < 0 \end{matrix}\right. 
\end{equation}
where $a$ and $c$ are the slopes for the power laws for stars dimmer (RGB stars) and brighter (AGB stars) than the TRGB, respectively, and $b$ dictates the RGB turnoff. The physical basis for this parameterization is the truncation of the red giant branch, which introduces a break in the stellar luminosity function.

In the literature, this break in the luminosity function is often found with a Sobel filter \citep{ogTRGB,sakai1996}. The Sobel filter is an algorithm that may be used to highlight edges. Since the TRGB corresponds with a sharp change in the luminosity function, a Sobel filter should be able to detect the TRGB with ease. The reason we choose to use a maximum likelihood algorithm is that the Sobel method becomes increasingly less accurate as the TRGB gets closer to the detection limit. By folding in our completeness and photometric errors, we are able to better define the shape of the luminosity function, and determine with less bias the true location of the TRGB magnitude. To avoid contamination from main-sequence stars and supergiants present in both CMDs, we limit our selections to stars with magnitudes between 24.9-26.6, and colors between 1.4-2.2.

\begin{figure*}
\figurenum{3}
\epsscale{1.0}
\plottwo{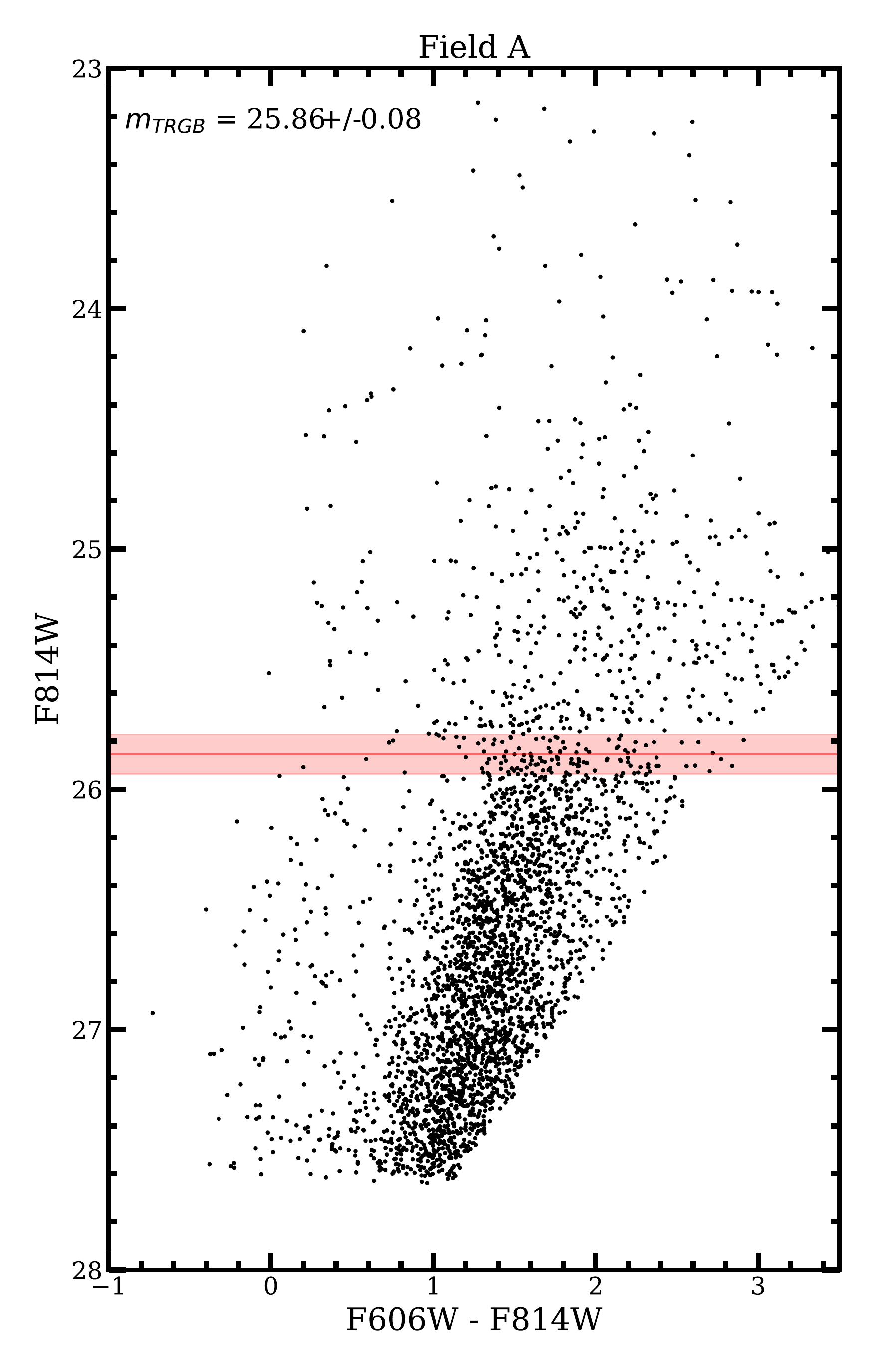}{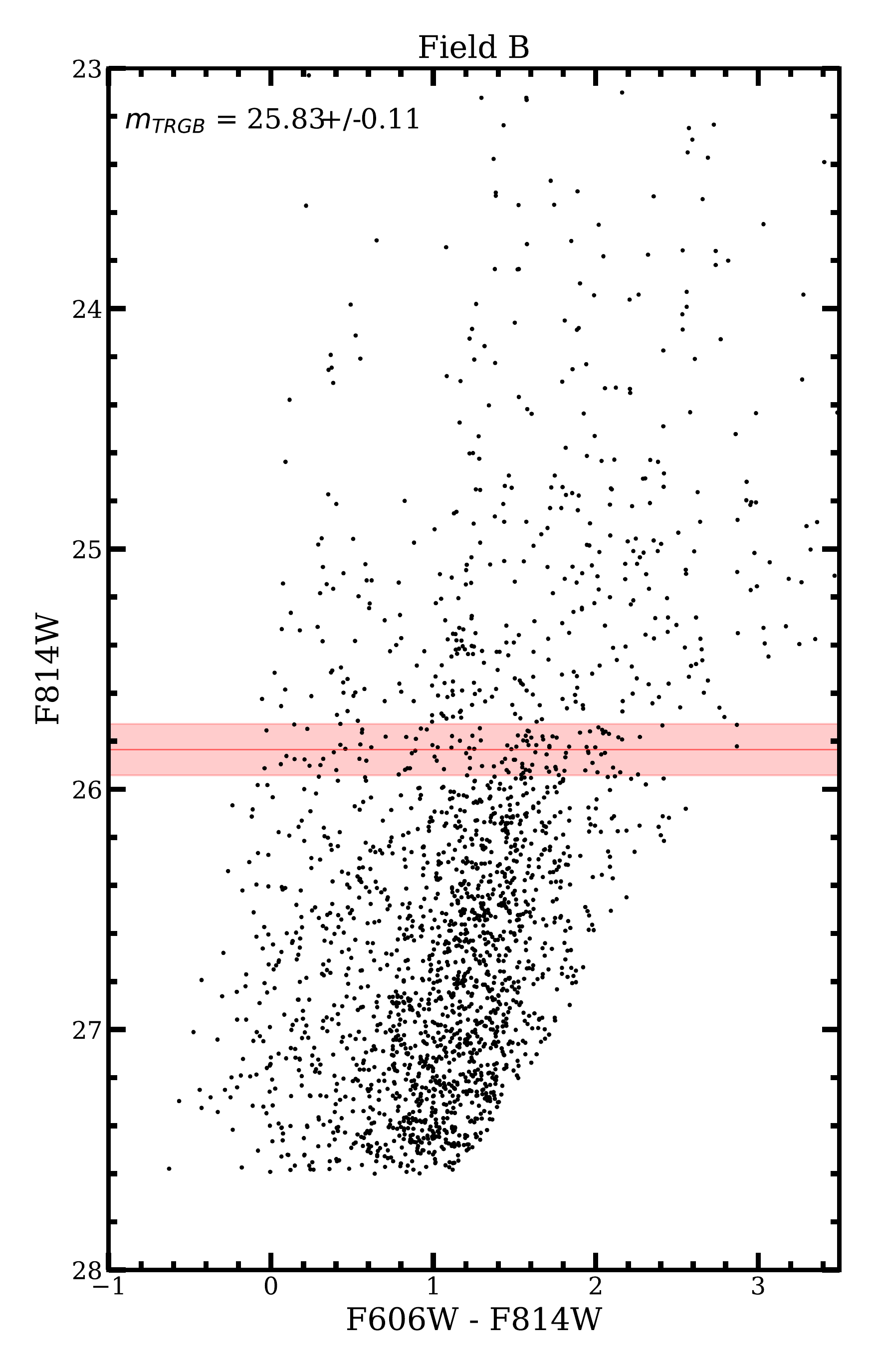}
\caption{This figure shows optical CMDs for  Fields A $\&$ B. The measured values for the apparent magnitude of the TRGB are shown by the red lines, where the light colored regions show 1$\sigma$ uncertainties. For both fields, we isolated regions with minimized Population I stars to reduce contamination. This technique was more successful for Field A, due to its position relative to the HI emission (Figure \ref{sidebyside}).}
\label{TRGBoptical}
\end{figure*}

The results for the optical fields are shown in Figure \ref{TRGBoptical}. We see that the apparent magnitude in the WFC3/UVIS system for the TRGB is $m_{TRGB}$ = 25.86 $\pm$ 0.08 for Field A, and $m_{TRGB}$ = 25.83 $\pm$ 0.11 for Field B. Note that these magnitudes are before accounting for the effects of dust along the line-of-sight, discussed in \S\ref{distance}. 

\begin{figure*}
\figurenum{4}
\epsscale{1.0}
\plottwo{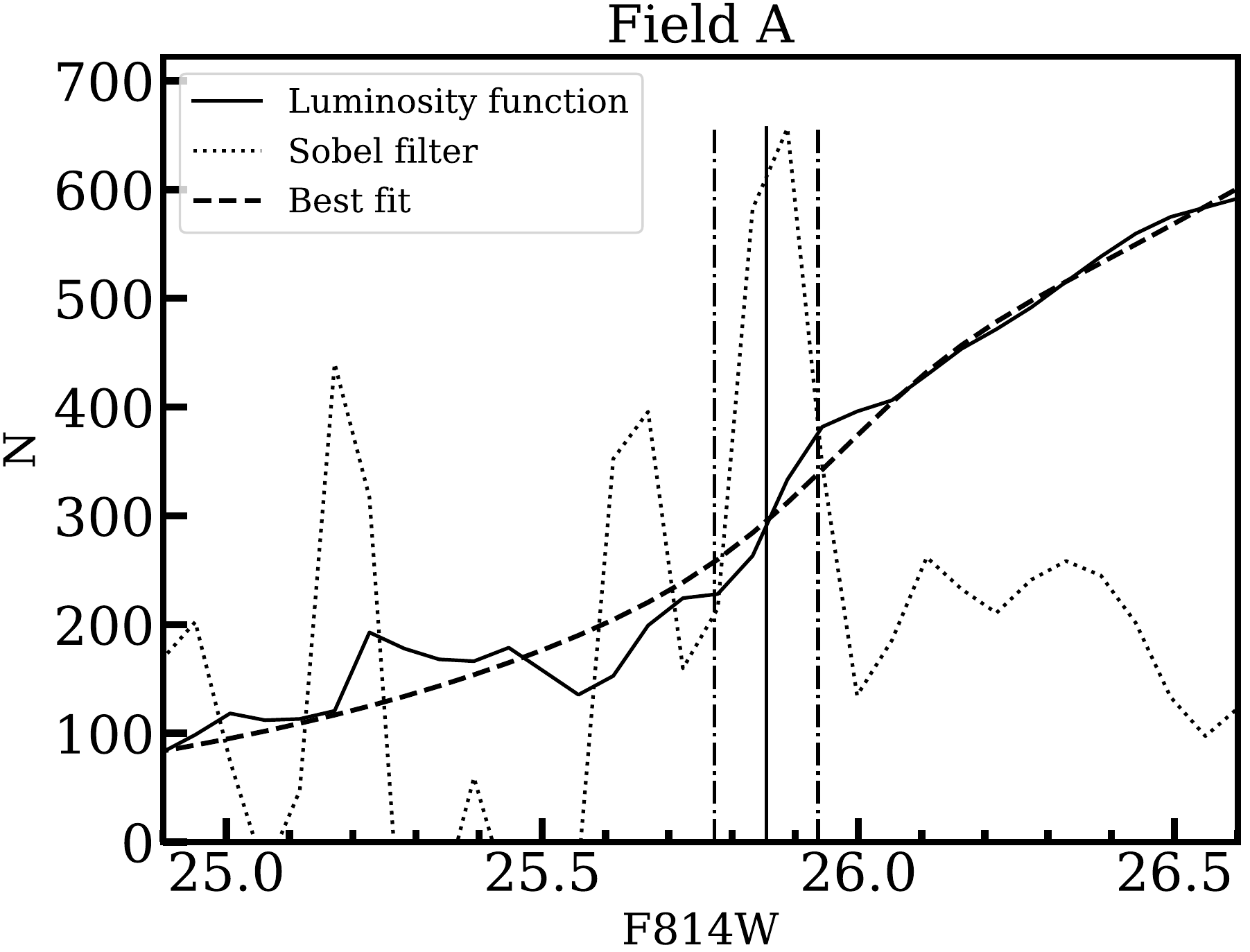}{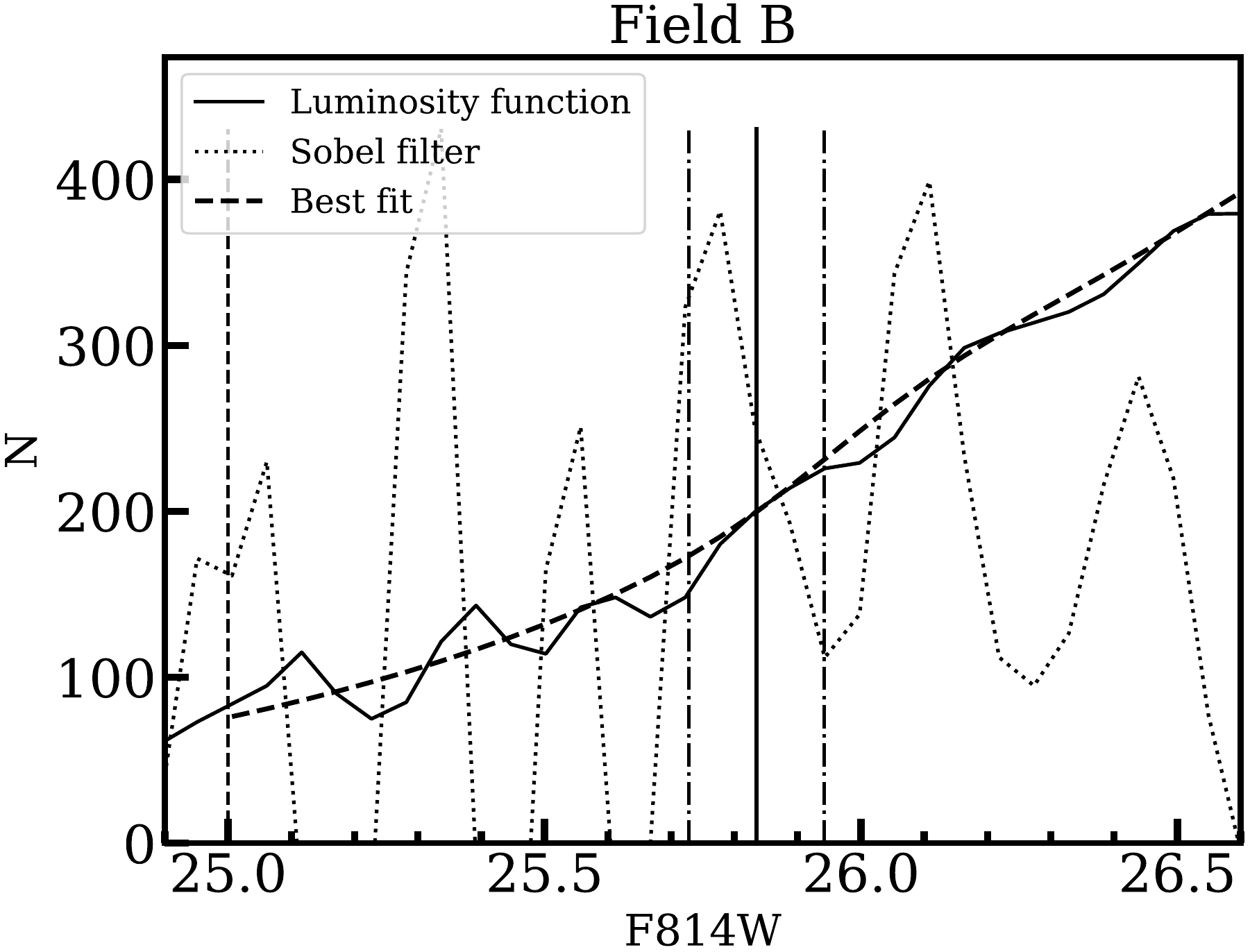}
\caption{This figure shows the results of fitting Equation 2 to our luminosity functions for each field. These figures also highlight results from a Sobel filtering analysis. We see that for Field A, the result matches with our maximum likelihood fit to within errors. For Field B, our TRGB magnitude lies between two Sobel peaks$-$ this result is likely due to a significant decrease in the number of stars as compared to Field A, which limits the usefulness of the Sobel filtering method.}
\label{lf}
\end{figure*}

We also show our fits to the luminosity functions for both fields in Figure \ref{lf}. We see that both fields have good fits. Both fields have similar best-fit parameters for $a$ ($\sim$0.3), $b$ ($\sim$0.1), and $c$ ($\sim$0.5). The plots also show the results of a Sobel filtering analysis on the number counts of galaxies, which we use as a check of our results. We find that for Field A, the result from the Sobel filter is within the maximum likelihood errors. For Field B, we find that the maximum likelihood TRGB magnitude lies between two Sobel peaks. This result is likely due to small number statistics, as the Sobel filter is extremely sensitive to changes in number counts.

After carrying through with the photometric reduction on the near-infrared fields, we find that they are simply not deep enough to locate the TRGB. 

\section{Distance Determination} \label{distance}

\subsection{Transformation to ACS Filters}
To be able to use existing TRGB calibrations \citep{rizzi2007} for Fields A $\&$ B, we need to transform our WFC3/UVIS magnitudes and colors to their corresponding ACS/WFC values. Although the filters themselves are close to identical, the cameras are not, leading to a small but non-negligible correction. The required transformations in our desired filters (F606W and F814W) were developed by \cite{janglee2015}, and are given by
\begin{equation}
\mathrm{F606W_{ACS}} = \mathrm{F606W_{WFC3}} + c_{0} - c_{1} (Color_{ACS})
\end{equation}
\begin{equation}
\mathrm{F814W_{ACS}} = \mathrm{F814W_{WFC3}} + c_{2} - c_{3} (Color_{ACS})
\end{equation}

where $Color_{ACS}$ = ($\mathrm{F606W_{ACS}}$ - $\mathrm{F814W_{ACS}}$), and 

\begin{equation}
\begin{array}{l}
c_{0} = 0.0016  \pm  0.0021 \\
c_{1} = 0.0322  \pm  0.0019 \\ 
c_{2} = 0.0156  \pm  0.0023 \\
c_{3} = 0.0060  \pm  0.0020 
\end{array}
\end{equation}

This set of equations allows us to perform all the required transformations, since we have photometry in both the F606W and F814W filters. These corrections are small ($\sim$0.01 mag in F814W, and $\sim$0.05 mag in F606W), as the photometric systems are very similar, as noted by \cite{janglee2015}.

\subsection{Extinction Corrections}
A source of error in a distance obtained from a TRGB measurement is the correction for dust extinction. One could use values obtained from all-sky dust maps \citep{SFD98, SF2011}. In our case, NGC 6946 is located at a low galactic latitude (b = 11.67$\degree$), which introduces a significant possibility for the extinction from these dust maps to be offset from the true value. To obtain a better estimate of the extinction in our fields, we use an approach similar to that of \cite{wu2014} and \cite{rizzi2017}. We employ a Sobel filter (with a bin size of 0.05) to measure the position of the zero-age main sequence (ZAMS) in color space. We then compare this value to that of NGC 300, a galaxy with very low extinction, to obtain a differential value for color excess. 

We know the color excess for NGC 300 to be E(B-V) = 0.01 \citep{SF2011}. We then use a value of $R_{V}$=3.1, the standard Fitzpatrick reddening curve \citep{F99}, and extinction coefficients $A_{filter}$/E(B-V) from \cite{SF2011} to convert the the color excess to total extinction in the appropriate \textit{HST} passbands. 

\begin{figure}
\figurenum{5}
\epsscale{1.1}
\plotone{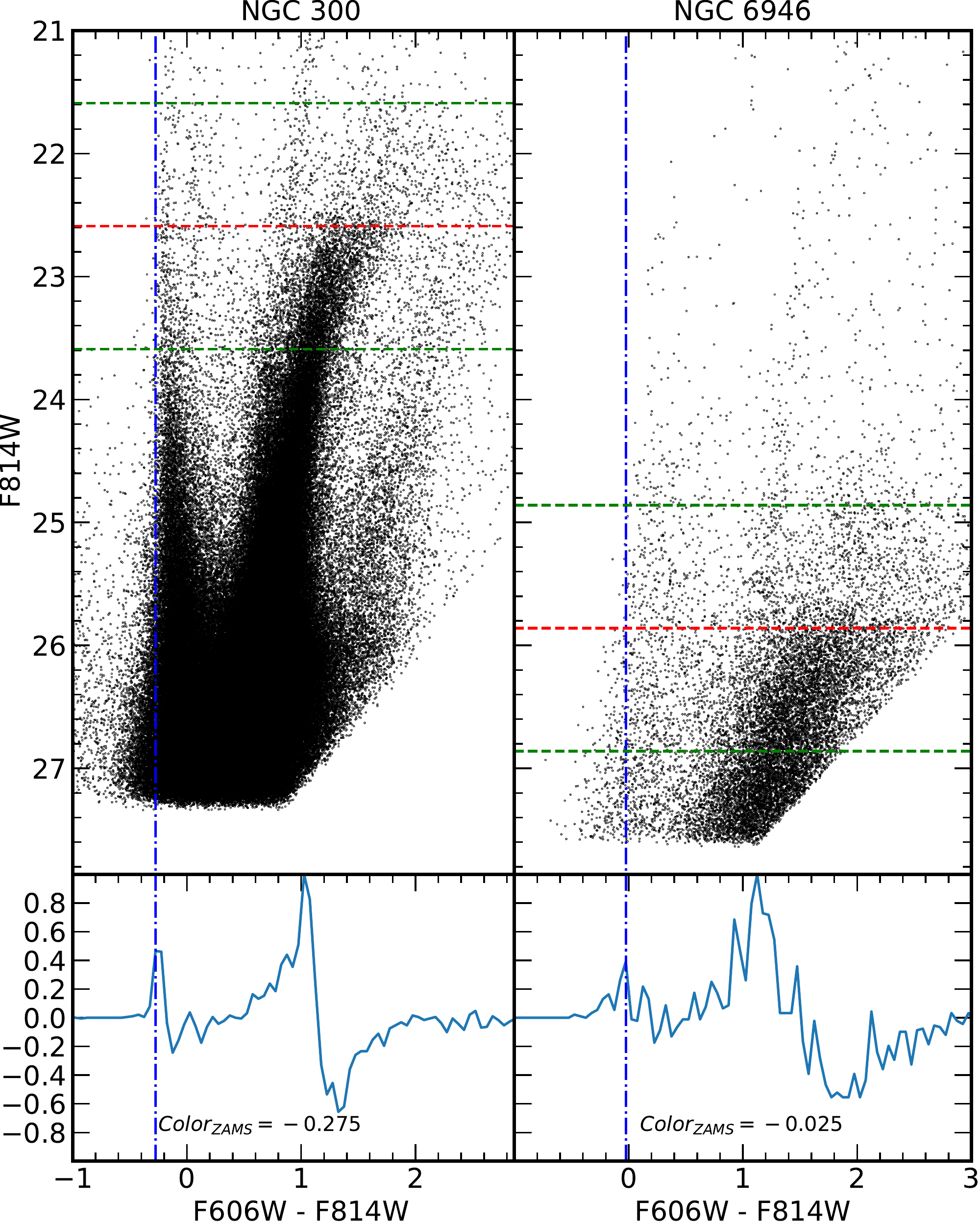}
\caption{This figure shows our measurement of the differential reddening between NGC 300 and NGC 6946 for the full Field A. The top panels are the CMDs for each galaxy, whereas the bottom are the measured responses of the Sobel filters for the stars that are within one magnitude (green) of the measured TRGB (red). The location of the zero-age main sequence is marked with the blue line.  Note that the CMD for NGC 300 is provided in the ACS/WFC system, whereas our observations of NGC 6946 are in the WFC3/UVIS system- we later converted the ZAMS color of NGC 6946 to the ACS/WFC system using the transformations provided by \cite{janglee2015}.}
\label{sobelOptical}
\end{figure}

For the fields observed with optical filters (Fields A $\&$ B), we obtained the relevant photometry for NGC 300 from the CMDs/TRGB catalog in EDD\footnote{\url{http://edd.ifa.hawaii.edu/}}, the Extragalactic Distance Database \citep{EDDCMD}. Figure \ref{sobelOptical} shows the results from our Sobel filtering analysis. We were not able to converge on a ZAMS color for Field B due to the fact that there are fewer stars in the field. Since the two fields are adjacent (Figure \ref{sidebyside}), we choose to adopt the extinction values from Field A for Field B as well. Since the photometry for NGC 300 is provided in the ACS/WFC system, we convert the NGC 6946 color values for Fields A and B out of the WFC3/UVIS system before direct comparison. After going through the appropriate transformations and comparing the two ZAMS colors from the pair of CMDs, we find $A_{F606W}$-$A_{F814W}$ = 0.27 $\pm$ 0.05. The estimate of the error includes contributions from the Sobel filtering, the original dust map \citep{SFD98} estimate for NGC 300, and the conversion between ACS/WFC $\&$ WFC3/UVIS.

\subsection{Absolute TRGB Magnitudes}

For the optical fields, we use the TRGB calibration provided by \cite{rizzi2007}. They find that the absolute magnitude of the TRGB has a slight dependence on color (and hence metallicity), and is given by:
\begin{equation}
M^{ACS}_{F814W} = -4.06 + 0.20[(\mathrm{F606W}-\mathrm{F814W})-1.23] 
\end{equation}
After converting the TRGB colors to the ACS/WFC system and correcting for dust, we find $M_{TRGB}$ = -4.03 for both Fields A ($\pm$ 0.03) $\&$ B ($\pm$ 0.04).

\subsection{Distances}

After applying all the above corrections and calibrations to our measured values, we are able to extract a final distance to NGC 6946. For the optical fields (A $\&$ B), we find distance moduli of 29.45 $\pm$ 0.12 and 29.43 $\pm$ 0.14, which correspond to distances of 7.74 $\pm$ 0.42 Mpc  and 7.69 $\pm$ 0.50 Mpc, respectively. The weighted average of these gives us a distance modulus of 29.44 $\pm$ 0.09, which corresponds to a distance of 7.72 $\pm$ 0.32 Mpc.

For comparison, we have also performed the above calculations assuming extinction values from \cite{SF2011}. We get distances of 7.51 $\pm$ 0.32 Mpc and 7.56 $\pm$ 0.41 Mpc for Fields A $\&$ B, which are slightly lower, but still consistent with the values we obtained by using extinctions obtained from our Sobel analysis.

A summary of our results is found in Table \ref{results}. In addition to this paper, we have uploaded the results (relevant values and color-magnitude diagrams) from this analysis to the Extragalactic Distance Database \citep{EDD, EDDCMD}.

\begin{table}[t]
\centering
\begin{tabular}{|l|c|c|}
\hline
                              & \textbf{Field A} & \textbf{Field B} \\ \hline
$\mathbf{m_{TRGB}}$           & 25.86 $\pm$ 0.08            & 25.84 $\pm$ 0.11     \\ \hline
\textbf{TRGB Color}           & 1.67 $\pm$ 0.08             & 1.64 $\pm$ 0.11      \\ \hline
\textbf{E(F606W-F814W)}       & 0.27 $\pm$ 0.05             & 0.27 $\pm$ 0.05      \\ \hline
$\mathbf{A_{814}}$            & 0.44 $\pm$ 0.08             & 0.44 $\pm$ 0.08      \\ \hline
$\mathbf{m_{TRGB,0}}$         & 25.42 $\pm$ 0.11            & 25.40 $\pm$ 0.13     \\ \hline
$\mathbf{(TRGB \ Color)_{0}}$ & 1.39 $\pm$ 0.13             & 1.36 $\pm$ 0.17       \\ \hline
$\mathbf{M_{TRGB,0}}$         & -4.03 $\pm$ 0.03            & -4.03 $\pm$ 0.04      \\ \hline
$\mathbf{(m-M)_{0}}$          & 29.45 $\pm$ 0.12            & 29.43 $\pm$ 0.14       \\ \hline
\textbf{Distance (Mpc)}       & 7.74 $\pm$ 0.42             & 7.69 $\pm$ 0.50       \\ \hline
\end{tabular}
\caption{This table shows the results from our TRGB analysis. Absolute TRGB calibrations, magnitudes and colors for all values are shown in ACS/WFC filters, transformed from the WFC3/UVIS flight system.}
\label{results}
\end{table}

\section{Discussion} \label{conc}
\subsection{Comparison to Previous Distances}

\begin{figure*}
\figurenum{6}
\epsscale{1.0}
\plotone{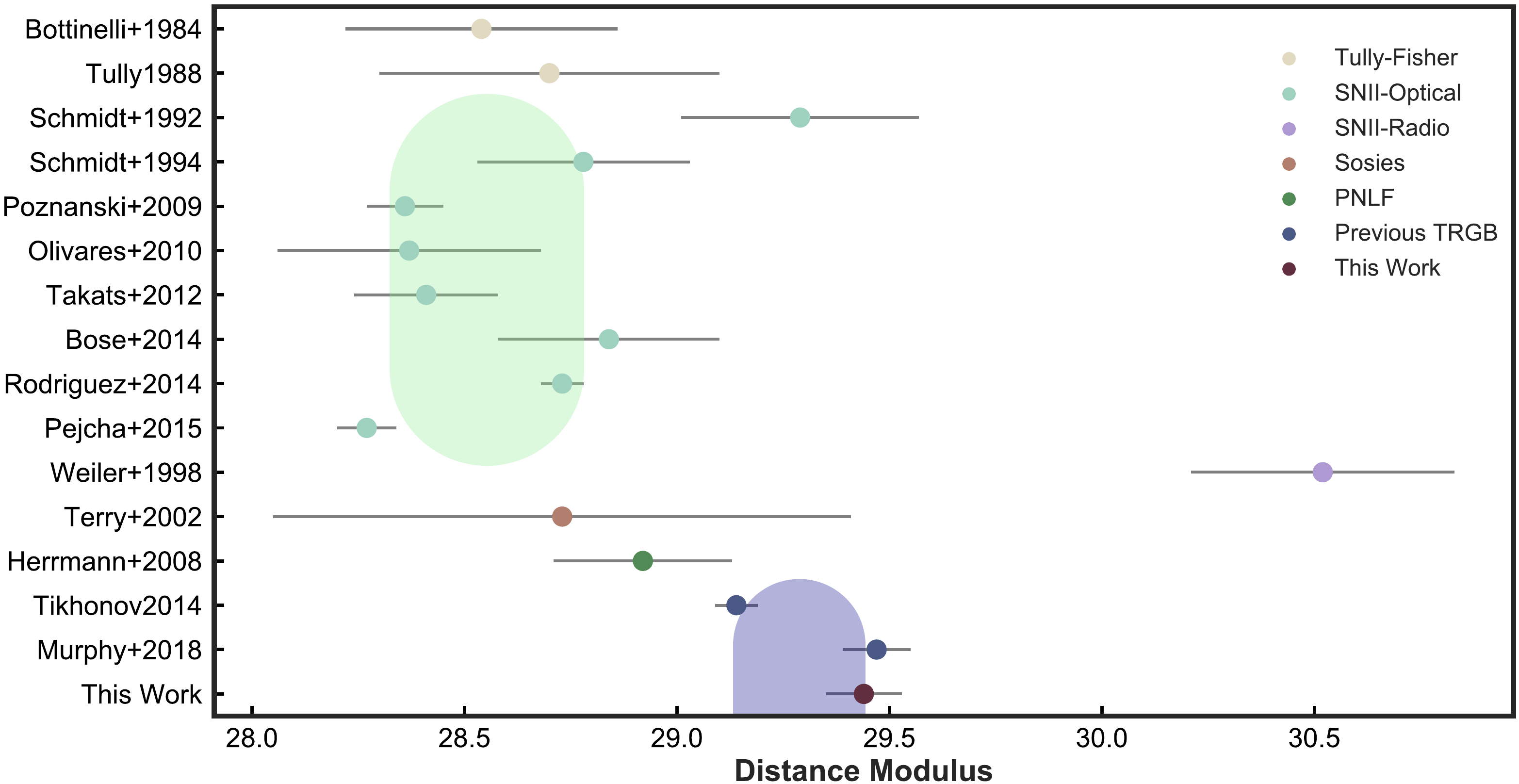}
\caption{A comparison of historical distance moduli to NGC 6946, color-coded by method used. For the supernovae and TRGB distances (including ours), we plot the weighted averages as vertical shaded regions, with the width determined by the standard deviation of the measurements. This highlights the large overall scatter with the type II supernova distances, despite the small reported errors with some of the individual measurements. Our value is systematically higher than the historical average, but closely matches the recent value from \cite{murphy2018}.}
\label{distancePlot}
\end{figure*}

In order to put our results into context, we plot previous determinations of distance moduli for NGC 6946 compared to ours (Figure \ref{distancePlot}). For the supernovae and TRGB distances, we plot the weighted averages as vertical shaded regions, with the width determined by the standard deviation of the measurements. This highlights the large scatter with the type II supernova distances, and encourages caution when using them as the associated errors are likely understated.

We obtained these distances from NED, excluding those without error measurements. For references with multiple distance measurements, we report the ones with the smallest associated errors. Distances obtained from the Tully-Fisher relation are likely to have large errors as the galaxy is too face-on for those measurements to be meaningful. Our weighted average distance modulus is 29.44 $\pm$ 0.09, which is higher than nearly all previous distances. Sources for distances not previously mentioned include \citet{B1984,NGC,S1992,S1994,radioDistance,T2002,P2009,O2009,T2012,B2014,R2014} and \cite{P2015}.

Our results can be compared to the other two TRGB distances in the literature. The TRGB distance from \cite{tikhonov} (d = 6.72 $\pm$ 0.15 Mpc) is lower than ours. Tikhonov performed his analysis on older ACS fields which are located much closer to the center of the galaxy, increasing the number of stars that contaminate the measurement. There is no statement of which calibration was used in the determination of the absolute magnitude of the TRGB. Additionally, the second band ACS observations he used were taken in F555W instead of F606W$-$ this choice led to the truncation of the red edge of the TRGB, which makes it difficult to determine its mean color. This ambiguity makes it difficult to calibrate the absolute magnitude of the TRGB, which contributes greatly to an uncertainty in his distance.

Our final distance is very close to that found in a recently published paper by \cite{murphy2018}, who found d = 7.83 $\pm$ 0.29 Mpc. To determine this distance, they use a similar maximum-likelihood modeling method described by \cite{mcquinn2016}. They performed their analysis on what we refer to as Field B throughout our paper, which is the field that has a higher contamination from the spiral arms of the galaxy. Unlike us, they do not perform spatial selections, leaving to even higher levels of possible contamination. There is also no mention of converting their WFC3/UVIS magnitudes to the ACS/WFC system before applying the calibrations of \citet{rizzi2007}. It is also worthwhile to note that their TRGB magnitude ($m_{TRGB}$ = 26.00 $\pm$ 0.04) is noticeably fainter than ours, while their estimate of the line-of-sight dust ($A_{F814W}$ = 0.57) is much greater. The source of their extinction value is unclear- they cite \citet{adamsDust}, who get their value of E(B-V) = 0.303 from \citet{SF2011}. This color excess corresponds to $A_{F814W}$ = 0.522 (ACS) or $A_{F814W}$ = 0.526 (WFC3), and not the reported $A_{F814W}$ = 0.57. For these reasons, we believe our value is the more robust one. Despite these reasons, the tension in their TRGB magnitude and extinction values with our results act in opposite directions upon distance, such that fortuitously our reported values agree within our respective error budgets.

The next most promising distance is one derived from the planetary nebula luminosity function (PNLF) by \citet{PNLF}. This technique can give distances to most galaxies within 10 Mpc, with uncertainties of $\sim$0.2-0.3 magnitudes, or 9-13$\%$ \citep{PNLFdistances}. However, in the case of NGC 6946, there are some complicating factors. Since the galaxy is a starburst, there are many HII regions, which \cite{PNLF} note can be mistaken for planetary nebulae. This potential confusion leads to a low sample size compared to the other galaxies in the study.

\subsection{Implications for Supernova Science}

Figure \ref{distancePlot} highlights the significant difference in distances obtained from type II supernovae compared to the TRGB values. The weighted mean distance modulus obtained from the supernovae is 28.55, with a standard deviation of 0.23. The resulting distance from supernovae is 5.13 Mpc, whereas our TRGB distance to the galaxy is 7.72 Mpc. This offset implies that the 10 supernovae detected in NGC 6946 are intrinsically brighter than previously thought. Assuming the TRGB distances, we find that the supernovae in NGC 6946 are on average $\sim$ 2.3 times more luminous than previous estimates. 

\subsection{Peculiar Velocity $\&$ the M101 Wall}
\begin{figure*}
\figurenum{7}
\epsscale{1.0}
\plotone{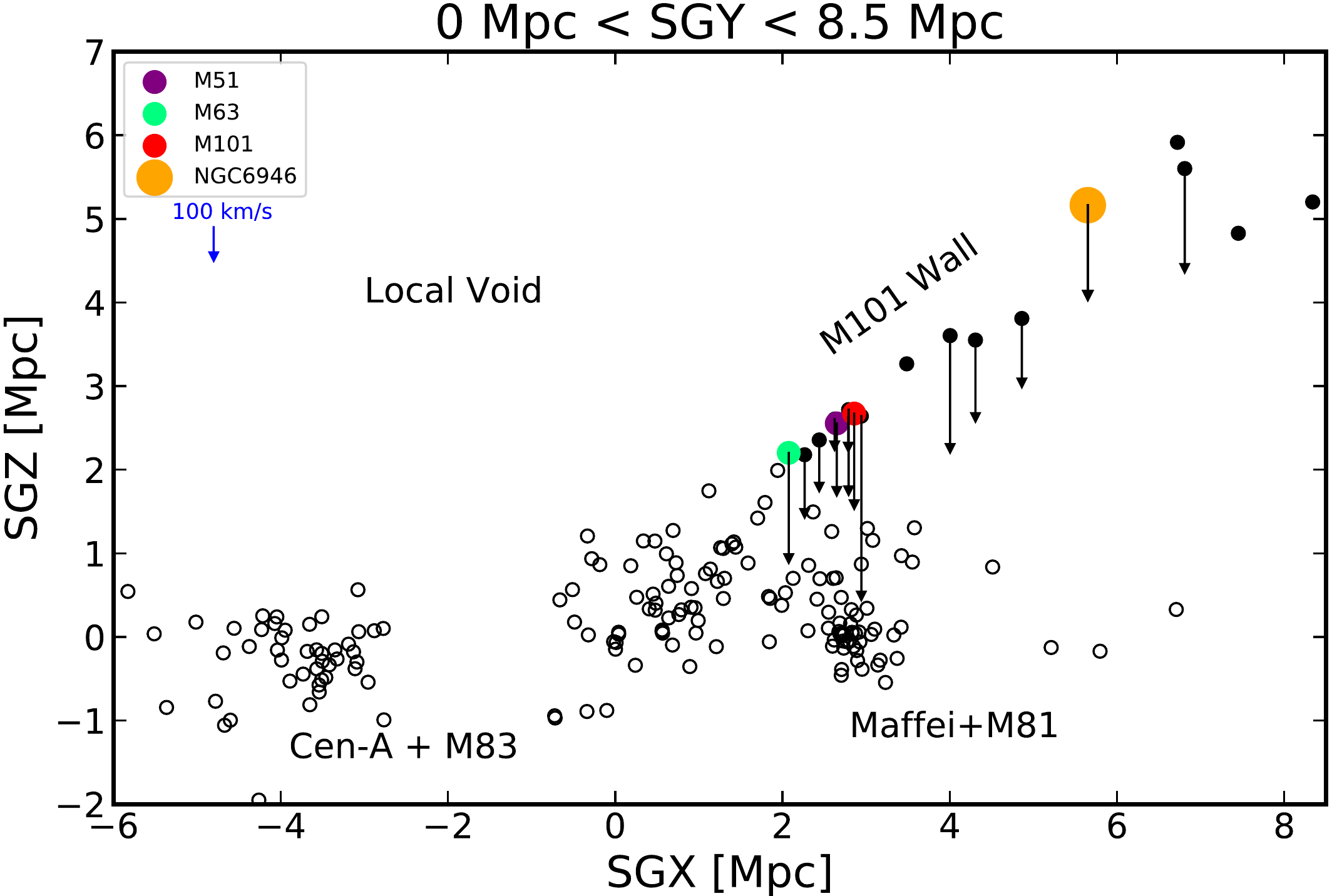}
\caption{This figure shows an SGX-SGZ projection for 0 Mpc $<$ SGY $<$ 8.5 Mpc. Open circles indicate galaxies with SGZ $<$ 2 Mpc, whereas filled circles indicate SGZ $\geq$ 2 Mpc. For galaxies with TRGB distances (and SGZ $\geq$ 2 Mpc), we show arrows indicating their SGZ motion.  Our distance to NGC 6946 is larger than most previous studies. The large absolute value of NGC 6946's peculiar SGZ motion indicates that it is rapidly evacuating away from the Local Void.}
\label{collapseSGY}
\end{figure*}

\begin{figure*}
\figurenum{8}
\epsscale{1.0}
\plotone{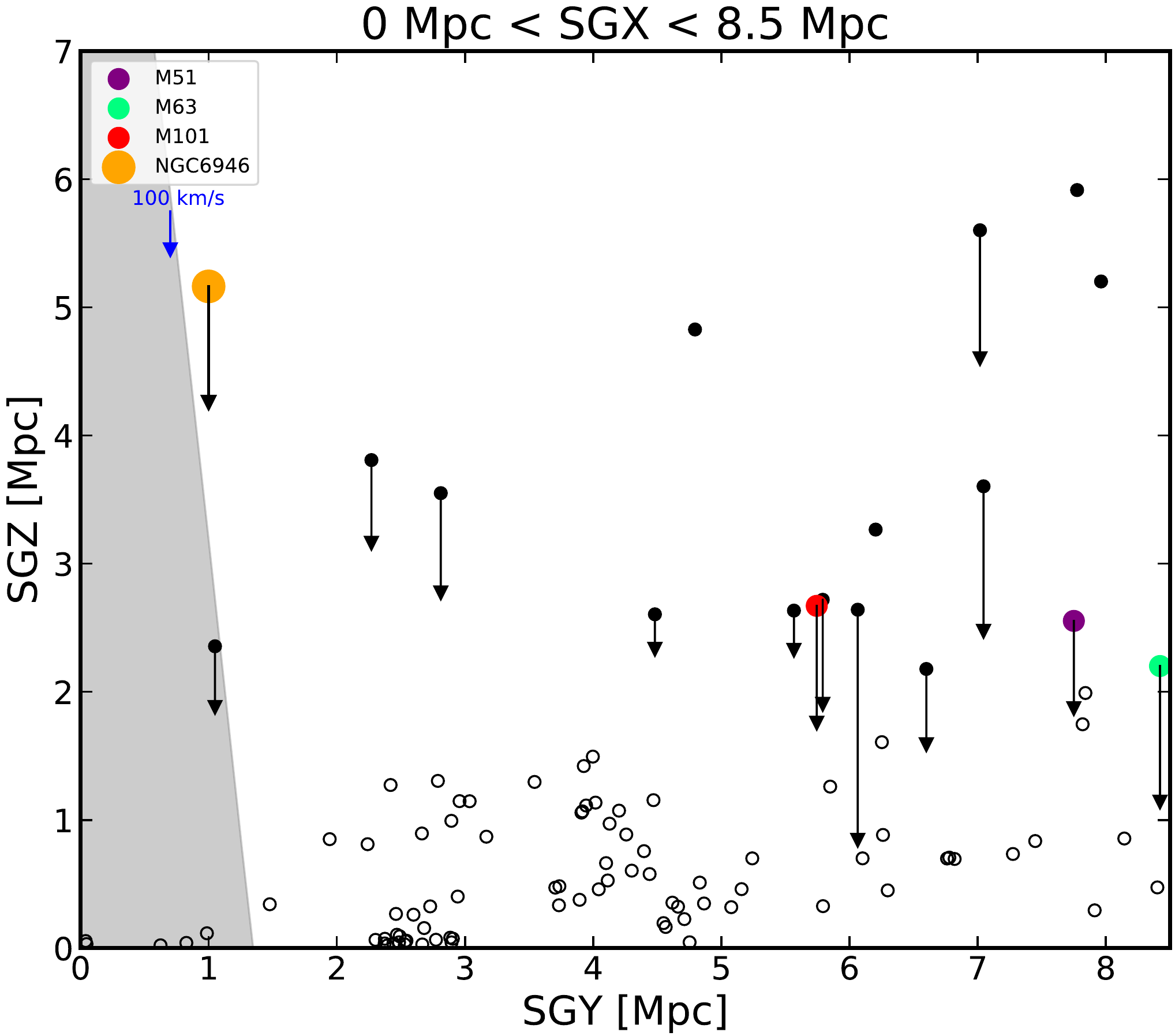}
\caption{This figure follows the same conventions as Figure \ref{collapseSGY}, but for an SGY-SGZ projection (for 0 Mpc $<$ SGX $<$ 8.5 Mpc). A projection of 10$^{\circ}$ from the galactic midplane at the distance of NGC 6946 is highlighted in gray. The coherent motions of the galaxies in the negative SGZ direction is clear.}
\label{collapseSGX}
\end{figure*}

NGC 6946 has an observed heliocentric velocity of 40 $\pm$ 2 km/s \citep{helio}. In the Local Sheet frame \citep{peculiar}, this transforms to a velocity of 350 $\pm$ 2 km/s. Solving Equation 1 and assuming $H_{0}$ = 75 $\pm$ 2 km/s/Mpc \citep{CF3}, we find a peculiar velocity in the Local Sheet frame of $v_{pec} = -229$ $\pm$ \textbf{$29$} km/s. This velocity is the projected component of a substantial negative SGZ motion ($v$ = $-342$ $\pm$ $43$ km/s), directed away from the Local Void. This motion is shown in both Figures \ref{collapseSGY} and \ref{collapseSGX}. 

Figure \ref{collapseSGY} shows an SGX-SGZ projection for 0 Mpc $<$ SGY $<$ 8.5 Mpc. Open circles indicate galaxies with SGZ $<$ 2 Mpc, whereas filled circles indicate SGZ $\geq$ 2 Mpc. For galaxies with TRGB distances (and SGZ $\geq$ 2 Mpc), we show arrows indicating their SGZ motion. Distances to galaxies with SGZ $\geq$ 2 Mpc but without TRGB distances are determined through the Tully-Fisher relation (or in one case, through surface brightness fluctuations). All the relevant values are taken from Cosmicflows-3 \citep{CF3}, and supplemented by results from the literature where needed \citep{mcquinn2016}. 

Four main structural features are highlighted by this projection. Two are the Centaurus A+M83 and Maffei+M81 complexes, which both lie in the Local Sheet with SGZ $\sim$ 0. The third feature is the M101 wall, which is a long filament that is visible nearly edge-on in this projection. This feature was found by \cite{muller2017} to have an \textit{rms} thickness of only  67 kpc, with a measured length of over 3 Mpc. Note that this analysis did not include NGC 6946, as a robust distance was not available at the time. The M101 wall includes several large galaxies, which are denoted with separate colored markers on the figure. With our distance to NGC 6946, we see that it lies along the thin wall, much higher off the supergalactic equator. The observed motion is the projected component of a substantial negative SGZ motion ($v$ = $-342$ $\pm$ $43$ km/s), indicating a coherent motion along with the other members of the M101 wall. The large absolute value of NGC 6946's SGZ motion indicates that it is rapidly evacuating away from the fourth main feature of the diagram, the Local Void. In this projection, the extent and emptiness of the Local Void is evident. It is becoming clear that the M101 wall is experiencing the expansion of the Local Void, as previously suggested by \citet{muller2017}. 

Figure \ref{collapseSGX} follows the same conventions as Figure \ref{collapseSGY}, but for an SGY-SGZ projection (for 0 Mpc $<$ SGX $<$ 8.5 Mpc). Additionally, a projection of $\pm$10$^{\circ}$ from the galactic midplane at the distance of NGC 6946 is highlighted in gray. The coherent motions of the galaxies in the negative SGZ direction is clear here as well.

In order to develop a more coherent picture of the dynamics of the M101 wall, it is necessary to obtain distances to more of its members. \cite{muller2017} found 15 dwarf galaxies  that may lie along the wall. There are also other large galaxies in this region that lack precise distances (such as NGC 5585). TRGB distances to these galaxies will help illuminate the nature of the M101 wall. Important questions still remain, including the likelihood of it manifesting the expansion of the Local Void, the behavior of the individual galaxy groups, and the overall distribution of galaxies along this filamentary structure.

\section{Conclusion}

We use two archival \textit{HST} fields of the outer regions of NGC 6946 from Proposal 14786 \citep{14786} to measure the F814W magnitude of the tip of the red giant branch. We apply the appropriate calibrations and corrections to the dataset, and find distances of 7.74 $\pm$ 0.42 Mpc and 7.69 $\pm$ 0.50 Mpc for the two fields. Finding the weighted average of the two gives an overall distance to the galaxy of 7.72 $\pm$ 0.32 Mpc, which is significantly higher than previous measurements from type II supernovae.  

The most recent TRGB distance to this galaxy was found by \cite{murphy2018}, who performed their analysis on one of the fields we used (Field B) and found d = 7.83 $\pm$ 0.29 Mpc. Though our distances match up well, there is tension in the underlying values. We measure brighter magnitudes for the TRGB than their single value (25.86 $\pm$ 0.08 and 25.83 $\pm$ 0.11, vs. 26.00 $\pm$ 0.04). This difference is somewhat offset by the fact that they use a larger value for extinction along the line of sight ($A_{F814W}$ = 0.44 vs $A_{F814W}$ = 0.57).

Our distance to NGC 6946 implies that the supernovae in this galaxy are on average $\sim$ 2.3 times more luminous than previous estimates. Our distance also enables us to calculate a peculiar velocity for NGC 6946. In the Local Sheet frame, we find a peculiar velocity of $v_{pec} = -229$ $\pm$ \textbf{$29$} km/s. This velocity is the projected component of a substantial negative SGZ motion ($v$ = $-342$ $\pm$ $43$ km/s), indicating a coherent motion along with the other members of the M101 wall. The large absolute value of this motion indicates that it is rapidly evacuating away from the Local Void. The origin of the M101 wall is likely the result of local coherence in the expansion of the Local Void, as previously suggested by \citet{muller2017}. TRGB distances for other galaxies that lie along the M101 wall will help clarify the full extent of the wall, as well as the dynamics of the individual groups.

\section{Acknowledgements}
G.A. would like to thank Andrew Dolphin for useful discussions regarding the data reduction process for IR photometry with DOLPHOT and Connor Auge, Ehsan Kourkchi, and Michael Tucker for useful discussions throughout the length of the paper. We also thank the anonymous referee for their helpful comments.

All of the data presented in this paper were obtained from the Mikulski Archive for Space Telescopes (MAST). STScI is operated by the Association of Universities for Research in Astronomy, Inc., under NASA contract NAS5-26555. 

Support for this work was provided by NASA through grant number HST-AR-14319 from the Space Telescope Science Institute, which is operated by AURA, Inc., under NASA contract NAS 5-26555.

\facility{\textit{HST} (WFC3)}

\end{document}